\documentclass[%
preprint,
showkeys,
 amsmath,amssymb,
 aps,
]{revtex4-2}

\usepackage{graphicx}
\usepackage{dcolumn}
\usepackage{bm}

\usepackage{xcolor}

\begin{document}

\preprint{APS/123-QED}

\title{Beyond the dipole approximation:\\
A compact operator form to describe magnetizable many-body systems}

\author{Dirk Romeis}
 \email{romeis@ipfdd.de}
\affiliation{Leibniz Institute of Polymer Research Dresden\\
Hohe Strasse 6, 01069 Dresden, Germany
}%


%

\date{\today}

\begin{abstract}
To describe the interactions in magnetically soft particle systems either numerical full-field methods or dipole models are used. Whereas the former are computationally challenging, simple dipole interactions are largely underestimating the actual forces when particles get closer. Based on the full 2-body solution, an analytic approximation scheme for many-body full-field interactions is developed. The concept is formulated in terms of an improved operator that is equivalent to the classical dipole form. The full interaction operator allows to describe cluster formation and dispersion among particles in applied magnetic fields very compactly and highly efficient. In view of its simple 'dipole-like' form, the implementation is straightforward in many areas where magnetically soft objects are used.
\end{abstract}

\maketitle



\begin{figure}[b]
\includegraphics[width=0.31\textwidth]{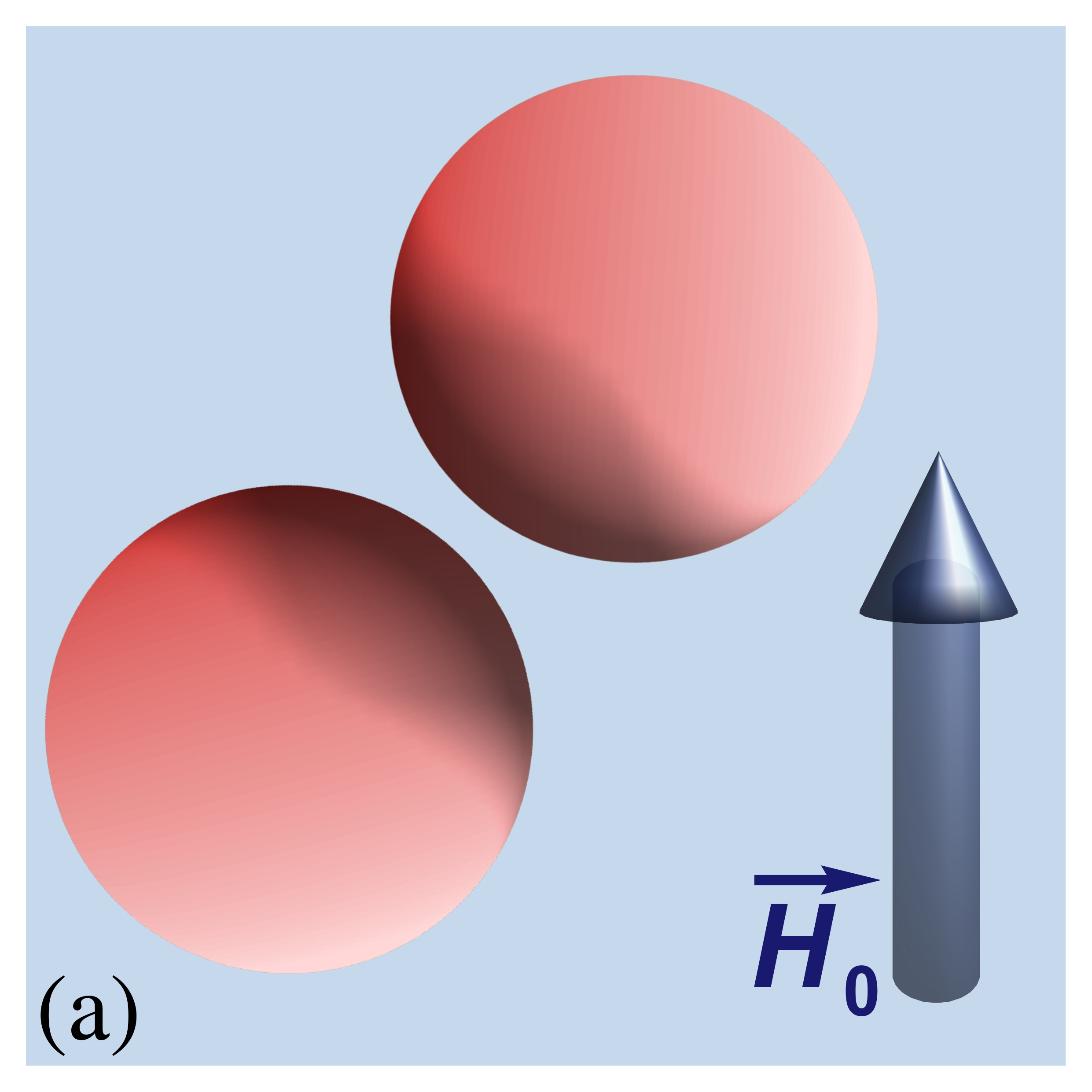}\includegraphics[width=0.31\textwidth]{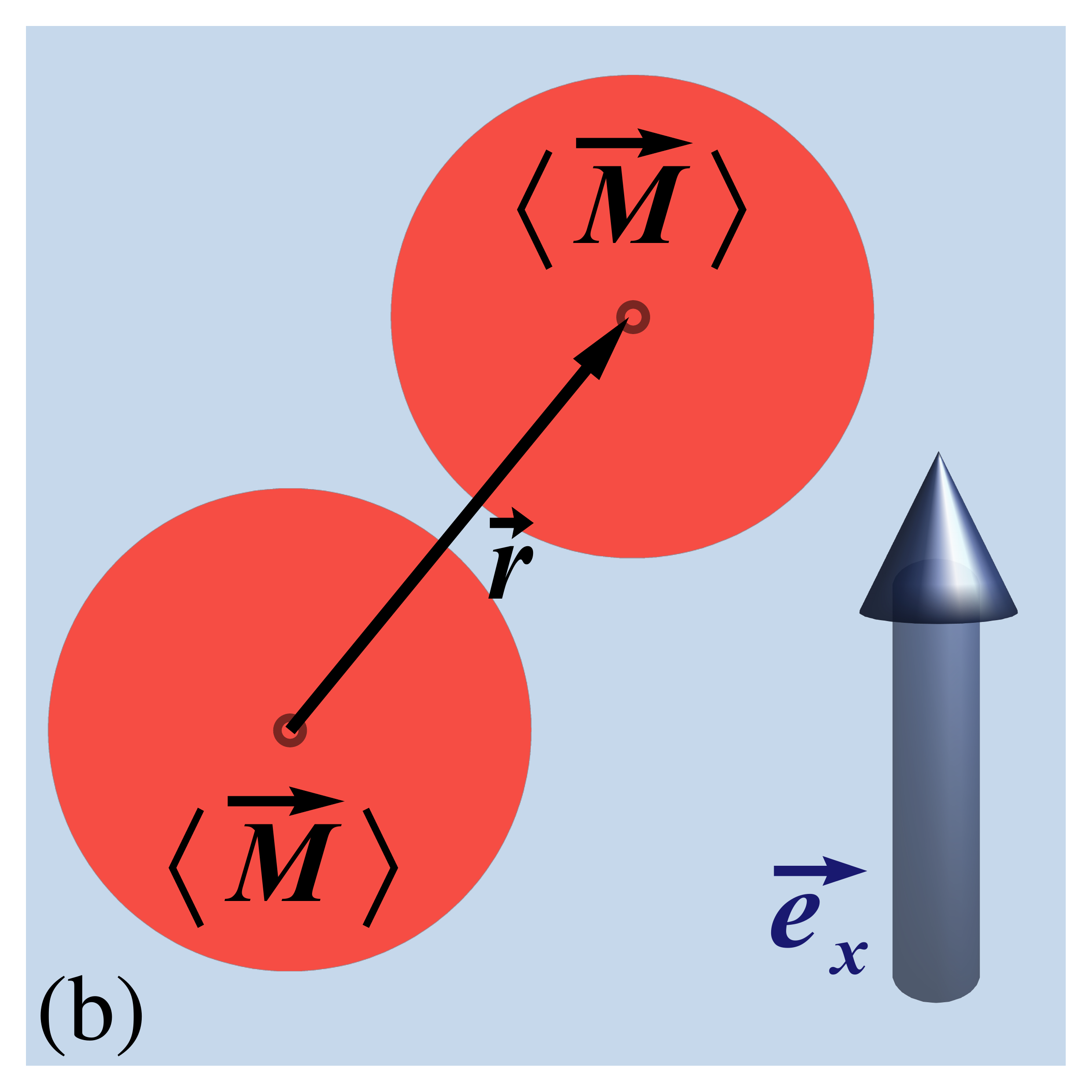}
\caption{\label{fig_2FF_vs_2MF} (a) Schematic visualization of magnetization inhomogeneities inside the volumes of two closely spaced spheres when an external magnetic field $\vec{H}_0$ is applied. The direction of $\vec{H}_0$ defines the $x$-axis. (b) Equivalent illustration in terms of the average formulation. The identical shade of red indicates that both spheres have the same average magnetization $\langle\vec{M}\rangle$. Vector $\vec{r}$ connects the centers of the spheres.}
\end{figure}

\begin{figure}[t]
\includegraphics[width=0.31\textwidth]{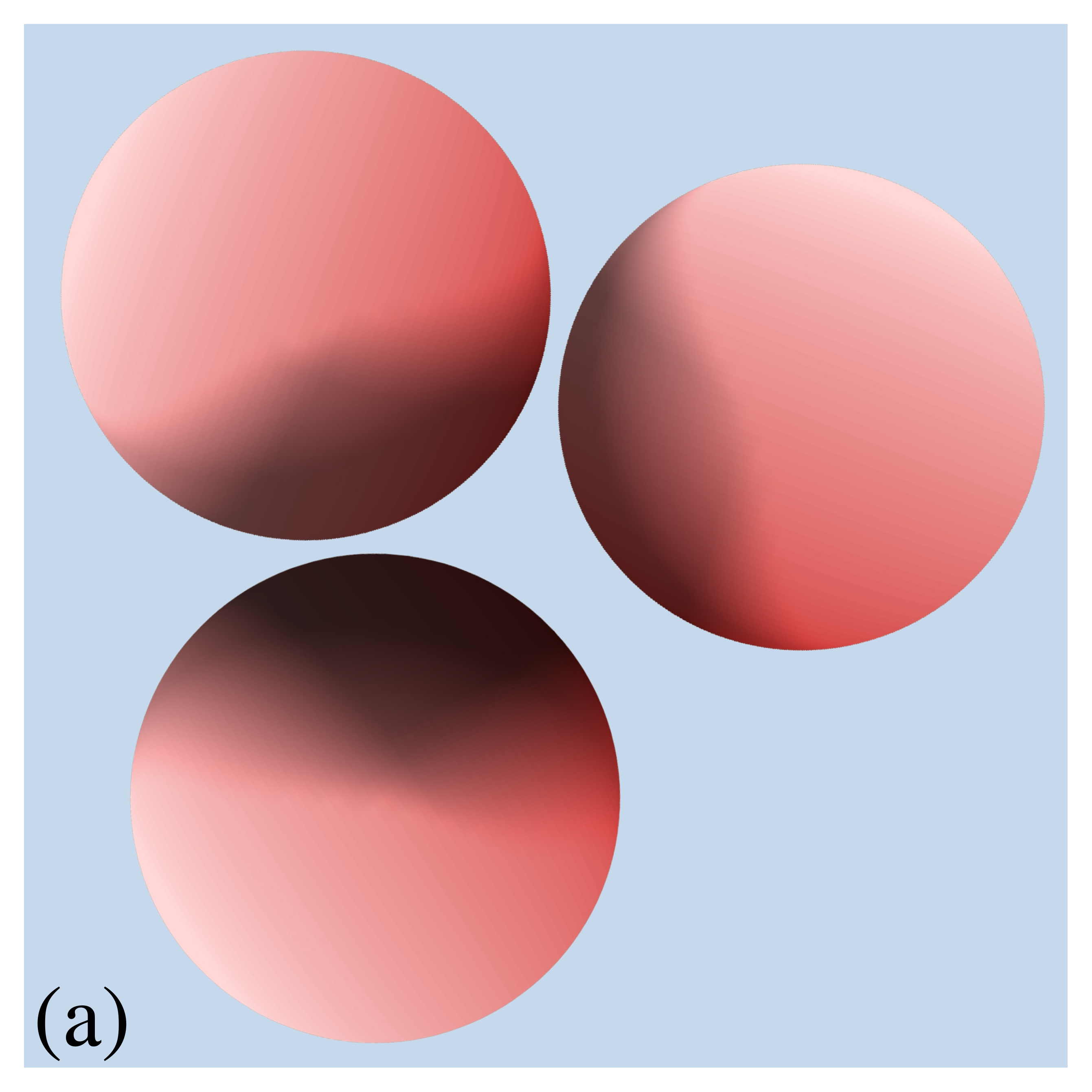}\includegraphics[width=0.31\textwidth]{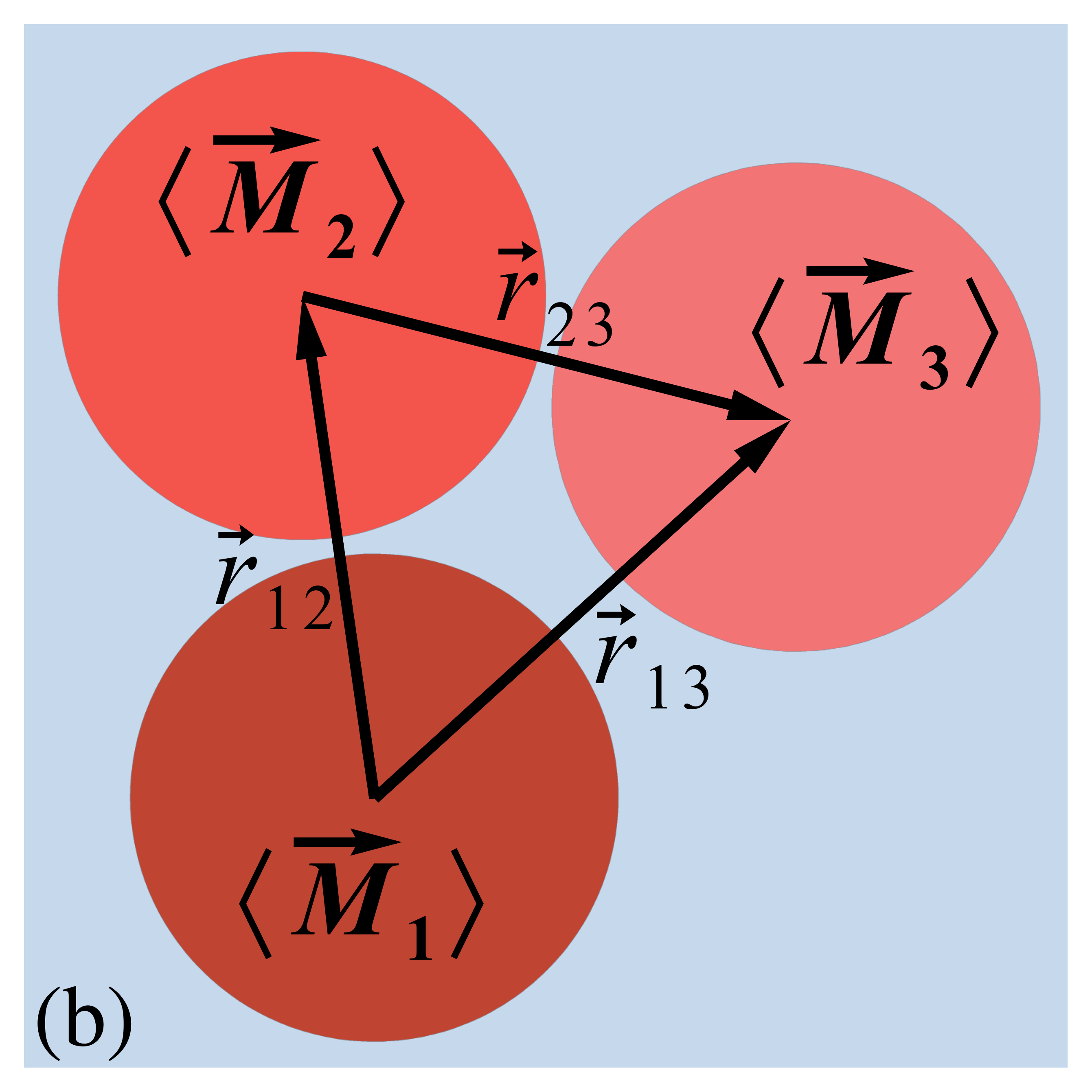}
\caption{\label{fig_FF_vs_MF} (a) Schematic visualization of magnetization inhomogeneities inside the volumes of three closely spaced spheres when an external magnetic field $\vec{H}_0$ is applied. (b) Equivalent illustration in terms of the average interaction operator. Different shades of red indicate that for each sphere a different self-consistent average magnetization $\langle\vec{M}_i\rangle$ is obtained.}
\end{figure}

{\it Introduction}\,{\bf --}\,Magnetic multidomain particles often exhibit paramagnetic behavior \cite{Nguyen2021, Silveyra2018, Filipcsei2007, Biller2014, BillerThesis, Biller2015, Puljiz2018}. Such soft magnetic particles are common in many areas of research and technologies, where they are used as a part of magnetorheological fluids, gels or elastomers \cite{Becker2019, Kostrov2021, Kubik2022}. They are also important in biomedical applications or soft robotics, as well as for manufacturing soft magnetic composite elements used in power electronics \cite{Attanayake2025, Miao2023, Silveyra2018}. Iron-based microparticles are typically employed for reasons such as cost-effective manufacturing, availability, excellent magnetic properties and biocompatibility. Furthermore, they are often of spherical shape \cite{Nguyen2021, Silveyra2018, Filipcsei2007} and, thus, most models consider magnetizable spheres \cite{Metsch2019, Mukherjee2020, Metsch2021, Lucarini2022, Romeis2016, Zubarev2016, Sanchez2019b, Isaev2019, Romeis2021, Goh2023, Stepanov2025, Fischer2026}. Due to their magnetization behavior, they are very sensitive to the local magnetic field. When two or more such particles approach, they noticeably influence each other. This is because the local magnetic field depends strongly on the exact arrangement of the particles \cite{Biller2014, BillerThesis, Biller2015}. In order to describe the interactions among them, a common way is to assign each particle a single point dipole located in the particle center. It is known that when multidomain particles are getting close to each other, the dipole approximation is no longer valid as it strongly underestimates the true interactions \cite{Biller2014, BillerThesis, Biller2015, Puljiz2018}. The discrepancy is caused by magnetization inhomogeneities inside each particle. This is schematically shown in Figs.~\ref{fig_2FF_vs_2MF}(a) and \ref{fig_FF_vs_MF}(a). Beyond dipole approximations, elaborate full-field methods \cite{Metsch2019, Mukherjee2020, Metsch2021, Lucarini2022} are used to capture these near-field effects. These more precise calculations are computationally very intensive. Already the simplest case of two linearly magnetizable spheres in close contact, requires hundreds of terms to achieve relative accuracy in the subpercentage range \cite{Biller2014, BillerThesis}. This impressively demonstrates the challenges to be faced in such many-body systems. Therefore, dipole approximations remain a common practical choice to model magnetic interactions in many-body systems \cite{Romeis2016, Zubarev2016, Sanchez2019b, Isaev2019, Romeis2021, Goh2023, Stepanov2025, Fischer2026}. Considering monodisperse spherical particles in the linear magnetization regime, and inspired by \cite{Yaremchuk2024}, a highly efficient alternative description is developed below. It is practically equivalent to the dipole formulation, but takes into account full-field interactions. 

{\it Self-consistent dipole description}\,{\bf --}\, By applying a magnetic field $\vec{H}_0$, a single sphere becomes homogeneously magnetized with magnetization $\vec{M}\parallel\vec{H}_0$. This creates an opposite demagnetization field of $-\vec{M}/3$ inside its volume \cite{Blundell_book}. Outside, a dipole field is induced, emanating from the center of the particle \cite{Blundell_book}. Linear magnetization behavior is typically described as $\vec{M}=\chi\vec{H}$, where $\chi$ is the magnetic susceptibility and $\vec{H}$ the total magnetic field \cite{Blundell_book}. If $N$ linearly magnetizable spheres are exposed to $\vec{H}_0$, the system of equations in the dipole approximation becomes \cite{Romeis2021}:
\begin{equation}
 \label{eq:DP_system}
 \vec{M}_i=\chi_{\text{eff}}\bigg(\vec{H}_0+\sum\limits_{\underset{j\neq i}{j=1}}^{N}\hat{g}(\vec{r}_{ij})\cdot\vec{M}_j\bigg).
\end{equation}
Here, $\chi_{\text{eff}}=\chi/(\chi+1/3)$ denotes the effective susceptibility \cite{Blundell_book, Romeis2021}, $\vec{M}_i$ the homogeneous magnetization of particle $i$, and the spatial vector $\vec{r}_{ij}$ connects the centers of particles $i$ and $j$. The central quantity is the dipole operator $\hat{g}$. In dimensionless form it reads:
\begin{equation}
 \label{eq:DP_operator}
 \hat{g}(\vec{r})=\frac{v_{\text{p}}}{4\pi}\left(-\frac{\hat{I}}{r^{3}}+\frac{3\vec{r}\vec{r}}{r^{5}}\right)=a^{3}\left(g_I(r)\hat{I}+g_R(r)\hat{R}\right).
\end{equation}
The volume of the sphere with radius $a$ is $v_{\text{p}}=4\pi a^{3}/3$. Furthermore, $\hat{I}$ denotes the identity tensor, and $\vec{r}\vec{r}$ the outer product. In the following, the particle radius is set to unity, i.e., $a=1$. On the right-hand side in Eq.~(\ref{eq:DP_operator}) we introduce the shorthand notation $\hat{R}:=\frac{\vec{r}\vec{r}}{r^{2}}$, and the coefficients are $g_I=-\frac{1}{3r^{3}}$ and $g_R=\frac{1}{r^{3}}$. The solution of Eq.~(\ref{eq:DP_system}) yields the self-consistent $\vec{M}_i$ for a given particle arrangement. With known $\vec{M}_i$ the total magnetic energy and the subsequent magnetic forces acting on each of the particles are determined straightforwardly \cite{Metsch2021}. In the case of just two interacting spheres, where due to symmetry $\vec{M}_1=\vec{M}_2=\vec{M}$, the equations reduce to:
\begin{equation}
 \label{eq:DP_2system}
 \vec{M}=\chi_{\text{eff}}\left(\vec{H}_0+\hat{g}(\vec{r})\cdot\vec{M}\right).
\end{equation}
Note, all fields, like $\vec{M}$ or $\vec{H}$, are considered homogeneous over the entire volume of each individual particle $i$. Thus, per definition they equivalently represent the volume average, i.e.,  $\vec{M}_i=\langle\vec{M}_i\rangle$ or $\vec{H}_i=\langle\vec{H}_i\rangle$. This assumption is valid if the spheres are not too close to each other \cite{Biller2014, BillerThesis, Biller2015}.

{\it Full-field solution for two interacting spheres}\,{\bf --}\,When two spheres in applied magnetic field come close to each other, pronounced inhomogeneities in the magnetization field appear. This is exemplified by the darker shading on the spheres in FIG.~\ref{fig_2FF_vs_2MF}(a). The exact solution for a system of two linearly magnetizable spheres
is obtained in terms of a slowly converging series expansion. Biller et al.~\cite{Biller2014, BillerThesis} developed a concise function that consists of just a few terms and reproduces the magnetic energy of the two linearly magnetizable spheres in applied field. 

In general, the magnetic energy of a system of linearly magnetizable particles in external field is given as \cite{Biller2015, Romeis2016}:
\begin{eqnarray}
 \label{eq:Umag}
U_{\text{mag}}&&=-\frac{\mu_0}{2}\int_{V_{\text{S}}}\!\!d^{3}r\sum_{i=1}^{N}\vec{M}_i(\vec{r})\cdot\vec{H}_0\nonumber\\
&&=-\frac{\mu_0}{2}H_0v_{\text{p}}\sum_{i=1}^{N}\langle M_{i\,x}\rangle\\
&&\label{eq:Umag2}\overset{N=2}{=}-\mu_0H_0v_{\text{p}}\langle M_{x}\rangle.
\end{eqnarray}
Here, $\mu_0$ is the vacuum permeability, and the integral in the first line is taken over the entire system volume $V_{\text{S}}$ where the applied field $\vec{H}_0=H_0\vec{e}_x$ is assumed to be homogeneous. Outside of the particles the material, e.g., polymer or carrier liquid, is non-magnetic. Thus, the integral over $V_{\text{S}}$ turns to $N$ separate integrals over the respective particle volumes. With homogeneous $\vec{H}_0$, each of these integrals yields the average magnetization $\langle\vec{M}_i\rangle$ in the respective particle times the volume $v_{\text{p}}$. The scalar product with $\vec{H}_0$ then returns the $x$-component $\langle M_{i\,x}\rangle$. Due to symmetry in the case $N=2$, see Eq.~(\ref{eq:Umag2}), the average magnetization in each particle must be the same, i.e., $\langle \vec{M}_1\rangle=\langle \vec{M}_2\rangle=\langle \vec{M}\rangle$. Comparing the function to resample the magnetic energy for two particles in \cite{Biller2014,BillerThesis} with the general form in Eq.~(\ref{eq:Umag2}), we identify: 
\begin{equation}
 \label{eq:Mx_2system}
\langle M_x\rangle=\chi_{\text{eff}}\,\mathcal{F}_{xx}H_0.
\end{equation}
Here, we introduced the solution factor $\mathcal{F}_{xx}=\mathcal{F}_{xx}(\vec{r})$. It yields the average magnetization in direction of $\vec{H}_0$. The form of $\mathcal{F}_{xx}$ is given in the End Matter in Table \ref{tab_paramet}. It essentially represents the function by Biller et al.~\cite{Biller2014,BillerThesis}. In the End Matter it is also shown that the complete vector $\langle \vec{M}\rangle$ is obtained from:
\begin{equation}
\label{eq:M_2system}
\langle \vec{M}\rangle=\chi_{\text{eff}}\,\hat{\mathcal{F}}\cdot\vec{H}_0\,,~~\text{with}~~\hat{\mathcal{F}}=\left(1+\mathcal{A}(r)\right)\hat{I}+\mathcal{C}(r)\hat{R}~.
\end{equation}
Here, we denote $\hat{\mathcal{F}}$ as the solution tensor. The scalar-valued functions $\mathcal{A}(r)$ and $\mathcal{C}(r)$ can be found in the End Matter in Table \ref{tab_paramet}. Reducing the full-field problem with inhomogeneities in $\vec{M}$, FIG.~\ref{fig_2FF_vs_2MF}(a), to the solution in form of Eq.~(\ref{eq:M_2system}) is graphically illustrated in FIG.~\ref{fig_2FF_vs_2MF}(b). 
 
\begin{figure}[t]
\includegraphics[width=0.58\textwidth]{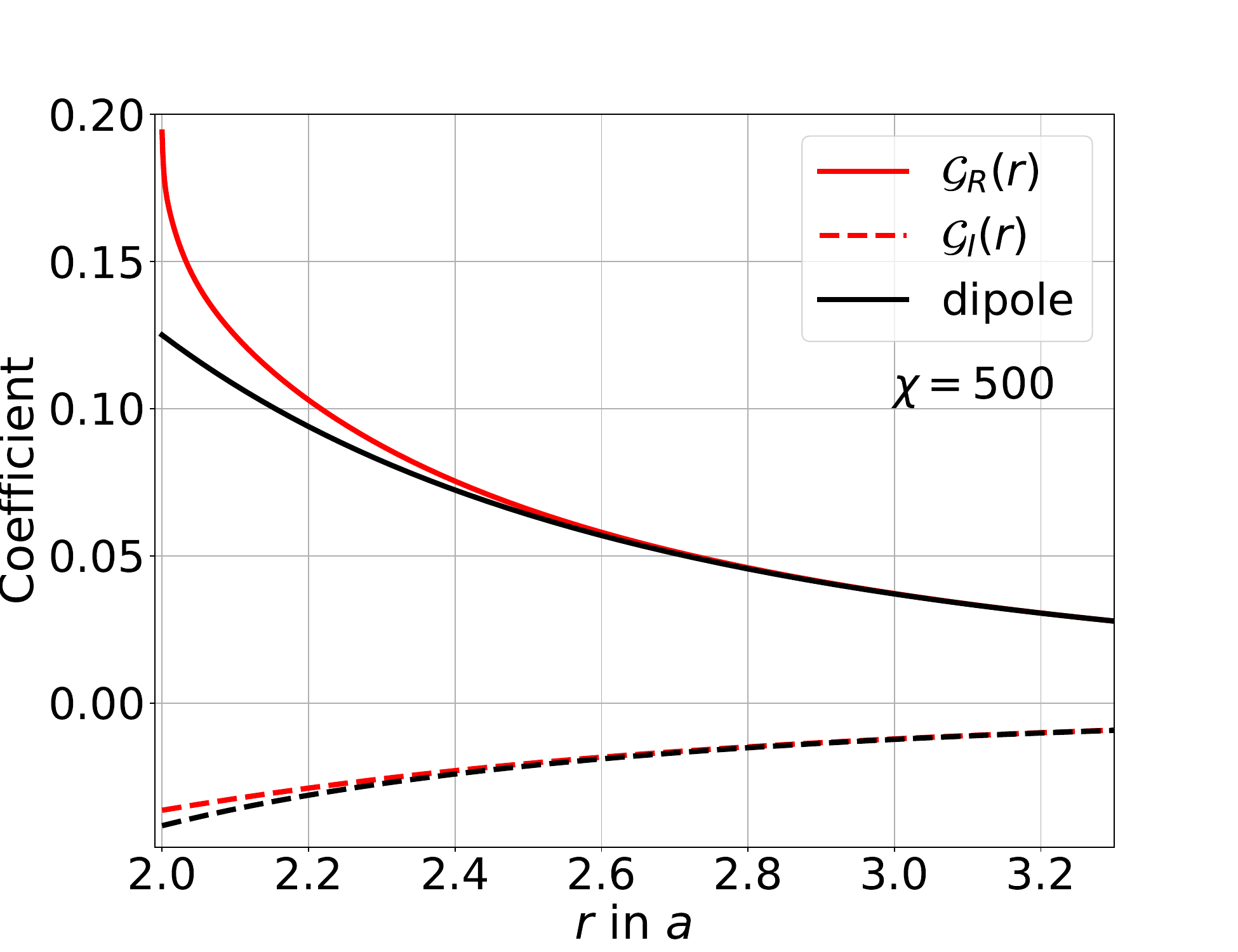}
\caption{\label{fig_Compare_Operators} Plot of the fundamental coeffiecents $\mathcal{G}_I(r)$ and $\mathcal{G}_R(r)$ (red colored) in dimensionless units, i.e., setting the sphere radius $a$ to unity. For comparison, also the equivalent dipole coefficients are plotted in black ($g_I=-\frac{1}{3r^{3}}$ and $g_R=\frac{1}{r^{3}}$).}
\end{figure}

{\it The average full-field interaction operator}\,{\bf --}\,In the dipole approximation, solving Eq.~(\ref{eq:DP_2system}) yields the self-consistent $\vec{M}$ for two interacting particles, i.e., dipoles. This solution can be written in equivalent form as Eq.~(\ref{eq:M_2system}) via a solution tensor $\hat{\mathcal{F}}^{\text{DP}}=\left(\hat{I}-\chi_{\text{eff}}\,\hat{g}\right)^{-1}$ for dipole interactions. In analogue form we now introduce an average full-field interaction operator $\hat{\mathcal{G}}$, which additionally takes into account near-field interactions. Equivalent to Eq.~(\ref{eq:DP_2system}), we write:
\begin{equation}
 \label{eq:FF_2system}
 \langle\vec{M}\rangle=\chi_{\text{eff}}\left(\vec{H}_0+\hat{\mathcal{G}}(\vec{r})\cdot \langle\vec{M}\rangle\right).
\end{equation}
Comparing Eqs.~(\ref{eq:DP_2system}) and (\ref{eq:M_2system}) one identifies:
\begin{equation}
 \label{eq:G_F_inversion}
\hat{\mathcal{G}}=\frac{\hat{I}-\hat{\mathcal{F}}^{-1}}{\chi_{\text{eff}}}.
\end{equation}
Due to the form of $\hat{\mathcal{F}}$, being the sum of the identity tensor $\hat{I}$ and the rank-one tensor $\hat{R}$, the inverse $\hat{\mathcal{F}}^{-1}$ can be computed analytically \cite{Miller1981} and $\hat{\mathcal{G}}$ takes the form:
\begin{equation}
 \label{eq:G_tensor}
\hat{\mathcal{G}}(\vec{r})=\mathcal{G}_I(r)\hat{I}+\mathcal{G}_R(r)\hat{R}.
\end{equation}
The scalar-valued functions $\mathcal{G}_I$ and $\mathcal{G}_R$ are listed in Table \ref{tab_paramet} and plotted in FIG.~\ref{fig_Compare_Operators} as functions of the interparticle
distance $r$. For a comparison against the classical dipole operator, see Eq.~(\ref{eq:DP_operator}), the corresponding dipole coefficients $g_I$ and $g_R$ are also shown in FIG.~\ref{fig_Compare_Operators}. As expected, with increasing particle distance $r$ the full-field coefficients approach to the dipole equivalents. The difference between full-field (red curves) and dipole (black curves) represent the direct near-field effects, which are clearly very short-range ($r\lesssim2.8a$). Nevertheless, considering a system with two particles in close contact and some other individual particles which are further away from the 2-particle cluster as well as from each other, the changes in the magnetization of the cluster particles, as altered by near-field effects, is indirectly mediated also to the distant particles via the long-range dipole interactions. These indirect near-field effects are therefore significantly more far-reaching than the direct effects. 

{\it Average full-field interactions in many-body systems}\,{\bf --}\,Analogue to the description of many-body interactions in the dipole approximation, as formulated in Eq.~\ref{eq:DP_system}, we now describe the full-field interactions via:
\begin{equation}
 \label{eq:FF_system}
 \langle\vec{M}_i\rangle=\chi_{\text{eff}}\bigg(\vec{H}_0+\sum\limits_{\underset{j\neq i}{j=1}}^{N}\hat{\mathcal{G}}(\vec{r}_{ij})\cdot\langle\vec{M}_j\rangle\bigg).
\end{equation}
Equivalent to the dipole approach, also here all mutual $i\leftrightarrow j$ interactions are first summed up. The solution provides the self-consistent $\langle\vec{M}_i\rangle$ where now additionally near-field interactions are taken into account. The massive reduction in computational costs and in mathematical complexity of the present operator form in comparison to the original full-field problem is schematically illustrated for a 3-particle system in FIG.~\ref{fig_FF_vs_MF}. Originally, the full-field problem
consists of calculating the inhomogeneous magnetization distribution as indicated in  FIG.~\ref{fig_FF_vs_MF}(a). Apparently, with increasing number of particles this becomes computationally intensive. In contrast, in terms of the elegant operator scheme illustrated in FIG.~\ref{fig_FF_vs_MF}(b), the full-field problem reduces to just 3 effective mutual interaction forms ($1\leftrightarrow2$, $1\leftrightarrow3$ and $2\leftrightarrow3$) which apply only to the center positions of the spheres. The self-consistent solution yields an approximation for the average magnetization in each sphere. The slightly different shades of red in FIG.~\ref{fig_FF_vs_MF}(b) are meant to symbolize different values for the average magnetization in each of the particles. Due to the considered linear magnetization behavior of the particle material, the here proposed linear superposition of mutual interactions in Eq.~(\ref{eq:FF_system}) is justified in good approximation. 

In most situations, when modeling many-body systems, the quantities of utmost interest are the forces acting on each of the particles due to the presence of the others. In the End Matter an analytic expression for the mutual magnetic forces is derived in close analogy to classical dipole forces. 

{\it Example calculations}\,{\bf --}\,In the following two small examples
\begin{figure}[b]
\includegraphics[width=0.58\textwidth]{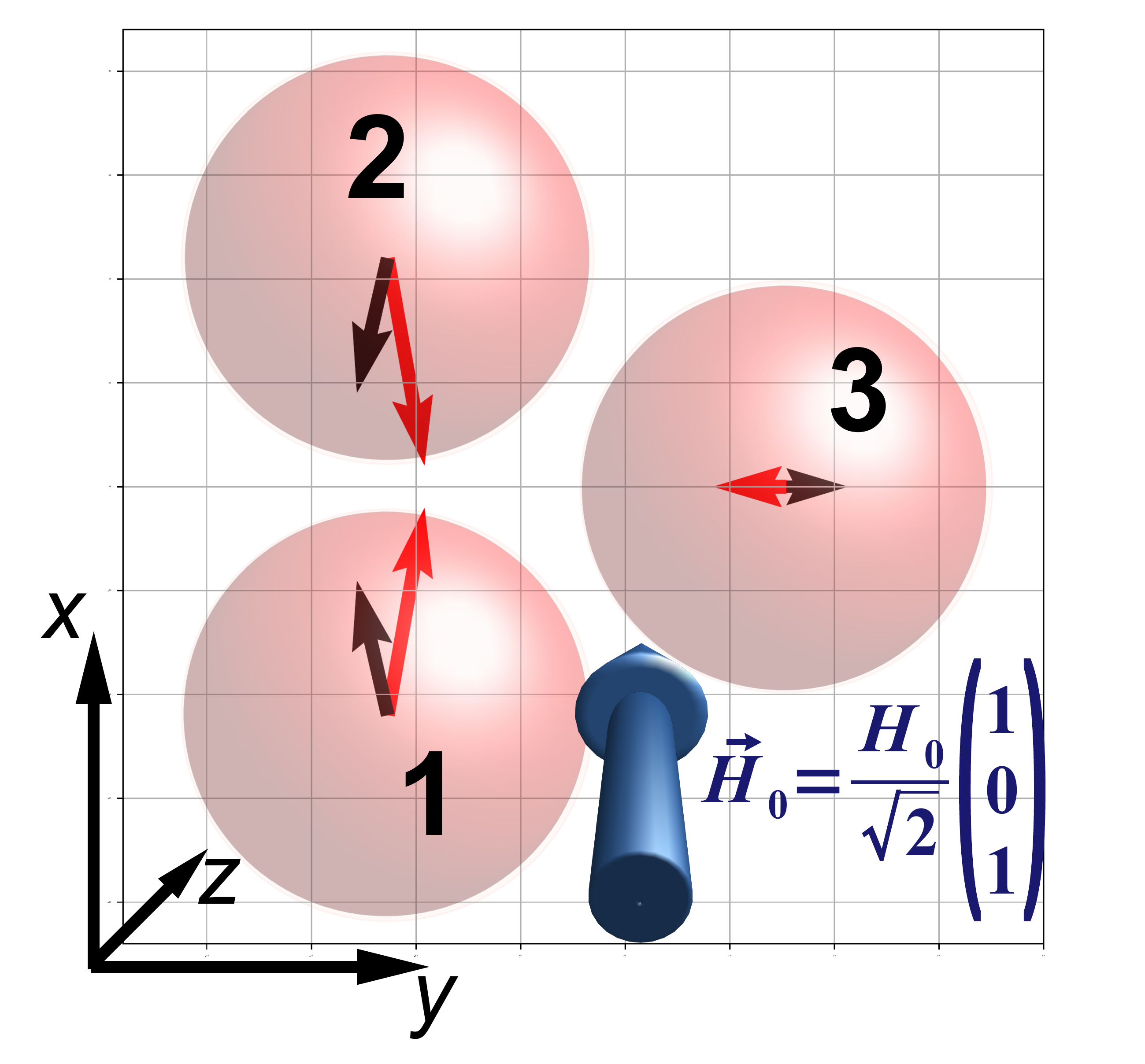}
\caption{\label{fig_3_sphere_example} First example: 3 spheres forming an equilateral triangle in the $x$-$y$-plane. A homogeneous field $\vec{H}_0$ is applied from diagonally above to this 3-particle system. The resulting body forces according to the self-consistent dipole (black arrows) and the self-consistent full-field approximation (red arrows) are drawn to the particle centers.}
\end{figure}
\begin{figure*}[t]
\includegraphics[width=0.93\textwidth]{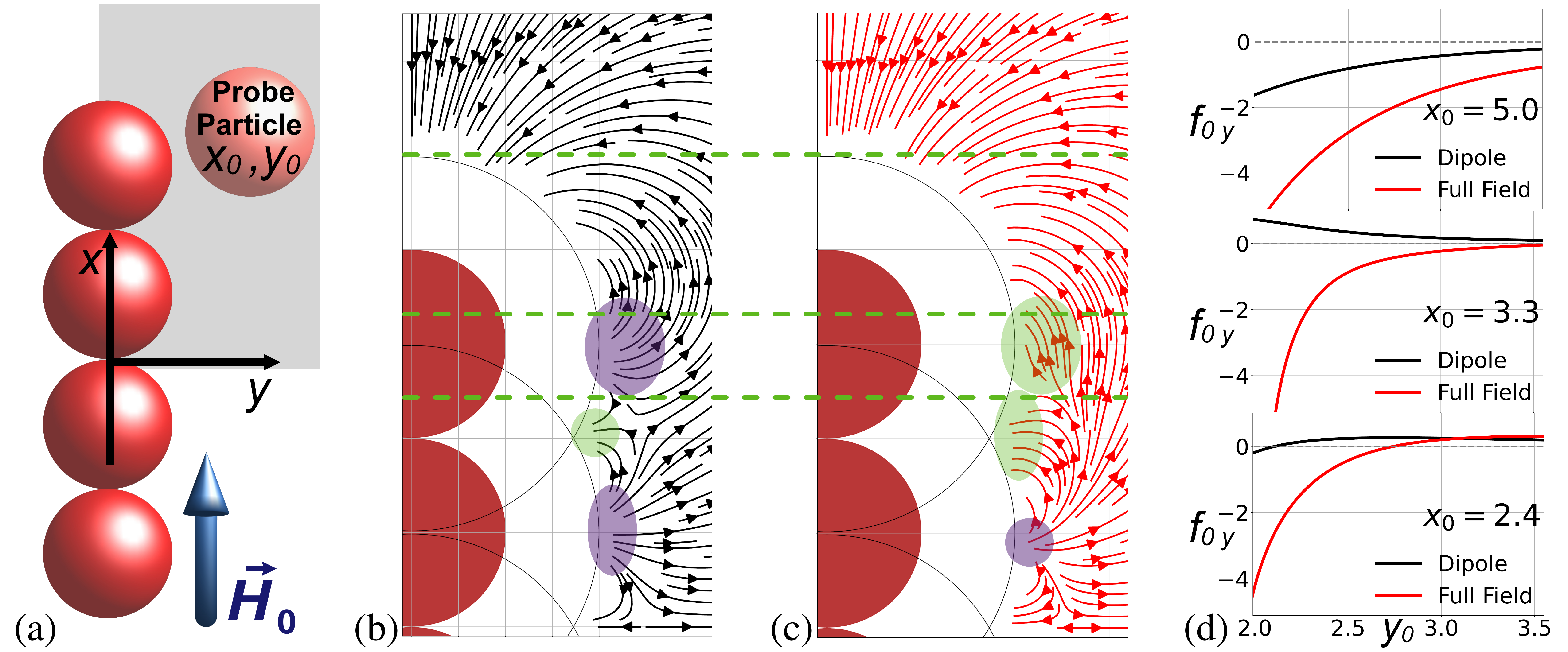}
\caption{\label{fig_5_sphere_example}(a) Second example: 4 spheres clustered into a chain along $\vec{H}_0$. An additional 'probe' particle is located somewhere arround. (b) The force field according to self-consistent dipole calculations in the form of streamlines showing the orientation of the body force that acts on the 'probe' particle located at that position. (c) Equivalent streamlines from self-consistent full-field calculations. (d) The lateral component $f_{0\,y}$ in $\mu_0 H_0^{2} a^{2}$ of the force $\vec{f}_0$ on the 'probe' particle at a fixed $x_0$ in dependence of its lateral distance $y_0$ from the chain cluster. The $x_0$-cuts are indicated by green dashed lines in (b) and (c).}
\end{figure*}
 are considered where the resulting forces according to the here developed full-field operator are compared against the corresponding predictions from the classical dipole operator. First, a system of 3 closely positioned spheres is examined. The center positions form an equilateral triangle, defining the $x$-$y$-plane, see FIG.~\ref{fig_3_sphere_example}. An external field $\vec{H}_0$ is applied from diagonally above. The resulting forces acting on the individual particles are indicated by vectors drawn to the centers of the three spheres. The impacts of the near-field effect are clearly evident when comparing the black arrows (dipole results) against the red ones (full-field results). While in the dipole approximation the sphere marked with the number 3 is repelled by the other two spheres, the results of the full-field approximation yield an attraction. Note that the strength of the forces is indicated by the length of the vectors, where the dipole forces are multiplied by a factor of 3 for better visibility. In true scales the dipole forces are significantly smaller than the full-field ones. Thus, including near-field effects, all 3 spheres experience strong attraction towards each other. In contrast, without near-field effects, sphere 3 would be repelled. 

In the second example we evaluate the forces exerted by a chain of 4 closely packed spheres on another 'probe' sphere, which is located at position $\vec{r}_0$ in the surroundings of this particle chain, see FIG.~\ref{fig_5_sphere_example}(a). That magnetizable particles form chain-like clusters in direction of the applied magnetic field is well-known from studies on magnetorheological fluids \cite{Zubarev2020, Lu2021} and magneto-active, a.k.a.~magnetorheological, elastomers \cite{Lu2021, Xu2019, Filipcsei2007}. The system investigated in FIG.~\ref{fig_5_sphere_example}(a) is rotational symmetric around $\vec{e}_x$, and it is symmetric about the $y$-axis. The relevant section is indicated by the gray shaded area. Except for regions forbidden by excluded volume, the 'probe' particle is located somewhere in the gray shaded area. The resulting forces $\vec{f}_0$ on this 'probe' particle are drawn in the form of streamlines with respect to its center position  $\vec{r}_0$ in FIG.~\ref{fig_5_sphere_example}(b) and FIG.~\ref{fig_5_sphere_example}(c). The black lines in FIG.~\ref{fig_5_sphere_example}(b) are the results for self-consistently interacting dipoles. The red lines in FIG.~\ref{fig_5_sphere_example}(c) indicate the forces according to self-consistent full-field calculations. Whereas the upper parts of FIG.~\ref{fig_5_sphere_example}(b) and (c) look very similar, the center and lower parts reveal clear differences. Here, the focus is on the indigo and green shaded areas of the streamlines. The indigo spots imply that there is a net repulsion, i.e., the force points away from the particle chain. In contrast, the green spots indicate a net attraction towards the chain. Clearly, full-field calculations predict significantly larger zones of attraction, and subsequent tendency that the single particle joins the chain, as compared to the dipole results. In FIG.~\ref{fig_5_sphere_example}(d) the component $f_{0\,y}$ of the force for a given $x_0$ position of the 'probe' particle is plotted in dependence of its lateral distance $y_0$. The three plots in FIG.~\ref{fig_5_sphere_example}(d) correspond to cuts through the force fields as indicated by the green dashed lines spanning FIG.~\ref{fig_5_sphere_example}(b) and (c). At $x_0=5.0$, although the streamlines are very similar, the magnitude of force predicted  differs strongly. Up to $y_0\sim3.5$ the attractive full-field forces (red) are almost 4 times larger than the dipole forces (black). This exemplifies the indirect impacts of near-field effects, as discussed before. At such distance direct effects are not relevant, see FIG.~\ref{fig_Compare_Operators}, but due to the close packing of the chain particles their magnetization is reinforced in the presence of near-field effects which results in substantial changes to the interactions with other particles that are much farther away. At $x_0=3.3$ in FIG.~\ref{fig_5_sphere_example}(d) dipole interactions predict exclusive repulsion at any $y_0$, but full-field calculations reveal broad attraction and a strong sink when $y_0<2.5$. Also at $y_0=2.4$ in FIG.~\ref{fig_5_sphere_example}(d) we note that while repulsive dipole forces nearly vanish, full-field forces are strongly attractive. 

{\it Conclusions}\,{\bf --}\,Based on the 2-body solution, an elegant analytical approach to describe magnetic many-body full-field interactions is developed. The concept is formulated in an equivalent way as known for the classical dipole model. Likewise, also the computational effort is comparable to dipole calculations, with the benefit that additional near-field effects are explicitly accounted for. The compact operator form allows to describe the process of clustering and dispersion between particles when modifying the applied field very efficiently and to significantly speed up the calculations. Such processes are considerably relevant in many research areas like magnetorheological fluids, gels or elastomers. Moreover, the here presented approach for magnetic multidomain interactions among monodisperse spherical particles features some general aspects, as expressed by the central Eq.~(\ref{eq:G_F_inversion}) of this work. Once the solution form for the 2-body system is obtained, an interaction operator can be formulated and applied to many-body systems. For example, a possible generalization to saturation magnetization could be envisioned: With increasing applied field, the magnetization 'hotspots', as suggested by dark red zones in FIG.~\ref{fig_FF_vs_MF}(a), become saturated first and will not be further magnetized by the presence of other nearby particles. Substantial changes in the local magnetization field then only occur in zones indicated by light pink color in FIG.~\ref{fig_FF_vs_MF}(a). Thus, linear superposition of mutual impacts is no longer a valid concept. Nevertheless, upon entering the saturation regime the magnetization distribution inside the particles also becomes increasingly homogeneous and the interactions return to the classical dipole model \cite{Biller2015, BillerThesis}. Hence, to extent the present scheme up to saturation one could employ an interpolation form with the full-field operator in Eq.~(\ref{eq:G_tensor}) at low magnetic fields and which gradually turns into the dipole operator in Eq.~(\ref{eq:DP_operator}) at saturation field. Alternatively, a solution form including saturation could be constructed and used in the inversion formula Eq.~(\ref{eq:G_F_inversion}). Generalizations towards systems of polydisperse spheres or differently shaped particles could be approached in an analogous way. In principle, the concept of turning the 2-body solution into an interaction formulation which then is applied in a self-consistent many-body description should be transferable not only to other magnetizable objects, but even to other kinds of physical interactions. 

{\it Acknowledgments}\,{\bf --}\,Financial support by Deutsche Forschungsgemeinschaft (DFG) to realize this research is gratefully acknowledged (No. RO 6756/3-1). The author would like to express his particular thanks to M. Saphiannikova for the very helpful discussions and useful suggestions during the course of this work.


\bibliography{references_NFE}

\appendix

\section{End Matter}

\begin{table}[b]
\caption{\label{tab_paramet}%
In the upper part, the numerical values of the parameters to reproduce the magnetic energy of two linearly magnetizable spheres in applied homogeneous magnetic field according to \cite{Biller2014,BillerThesis} are listed. In the lower part, the explicit form of the function $\mathcal{F}$, as developed in \cite{Biller2014,BillerThesis} and adopted to the present notation, is given. Additionally, coefficients $\mathcal{G}_I$ and $\mathcal{G}_R$ of the average interaction operator $\hat{\mathcal{G}}$ are presented.
}
\begin{ruledtabular}
\begin{tabular}{cccccc}
\multicolumn{6}{c}{\textrm{Parameters}}\\[0.2em]
$k$& \textrm{$A_k$} & \textrm{$B_k$} & \textrm{$C_k$} & \textrm{$D_k$} & \textrm{$p_k$}\\
\colrule
1 & -1 & 0 & 3 & 0 & 2\\
2 & 0 & 0 & $3.42\times10^{-2}$ & 1.3976 & 3\\
3 & 0.111 & -0.689 & $2.83\times10^{-6}$ & 1.8947 & 11\\
4 & 0.509 & 0.589 & $1.8\times10^{-13}$ & 1.9898 & 13\\
5 & -0.424 & 0.592 & 0 & 0 & 20\\
\hline\\[-0.4em]
\multicolumn{6}{c}{\textrm{Functional forms}}\\[0.2em]
\multicolumn{6}{c}{\(\displaystyle {\mathcal A}(r)=\sum\limits_{k=1}^{5}\, \left(\frac{\chi_{\text{eff}}}{3}\right)^{p_k-1}\frac{A_k}{(r-B_k)^{k+2}}\)}\\
\multicolumn{6}{c}{\(\displaystyle{\mathcal C}(r)=\sum\limits_{k=1}^{5}\,\left(\frac{\chi_{\text{eff}}}{3}\right)^{p_k-1}\frac{C_k}{(r-D_k)^{k+2}}\)}\\[1em]
\multicolumn{6}{c}{\(\displaystyle\mathcal{F}(\vec{r})= 1 + {\mathcal A}(r) + {\mathcal C}(r)\frac{(\vec{r}\cdot\vec{e}_x)^{2}}{r^{2}}\)}\\[1em]
\multicolumn{6}{c}{\(\displaystyle\mathcal{G}_I(r)=\frac{{\mathcal A}(r)}{\chi_{\text{eff}}\big(1+{\mathcal A}(r)\big)}\)}\\[1em]
\multicolumn{6}{c}{\(\displaystyle\mathcal{G}_R(r)=\frac{{\mathcal C}(r)}{\chi_{\text{eff}}\big(1+{\mathcal A}(r)\big)\big(1+{\mathcal A}(r)+{\mathcal C}(r)\big)}\)}\\
\end{tabular}
\end{ruledtabular}
\end{table}

\begin{figure}[t]
\includegraphics[width=0.58\textwidth]{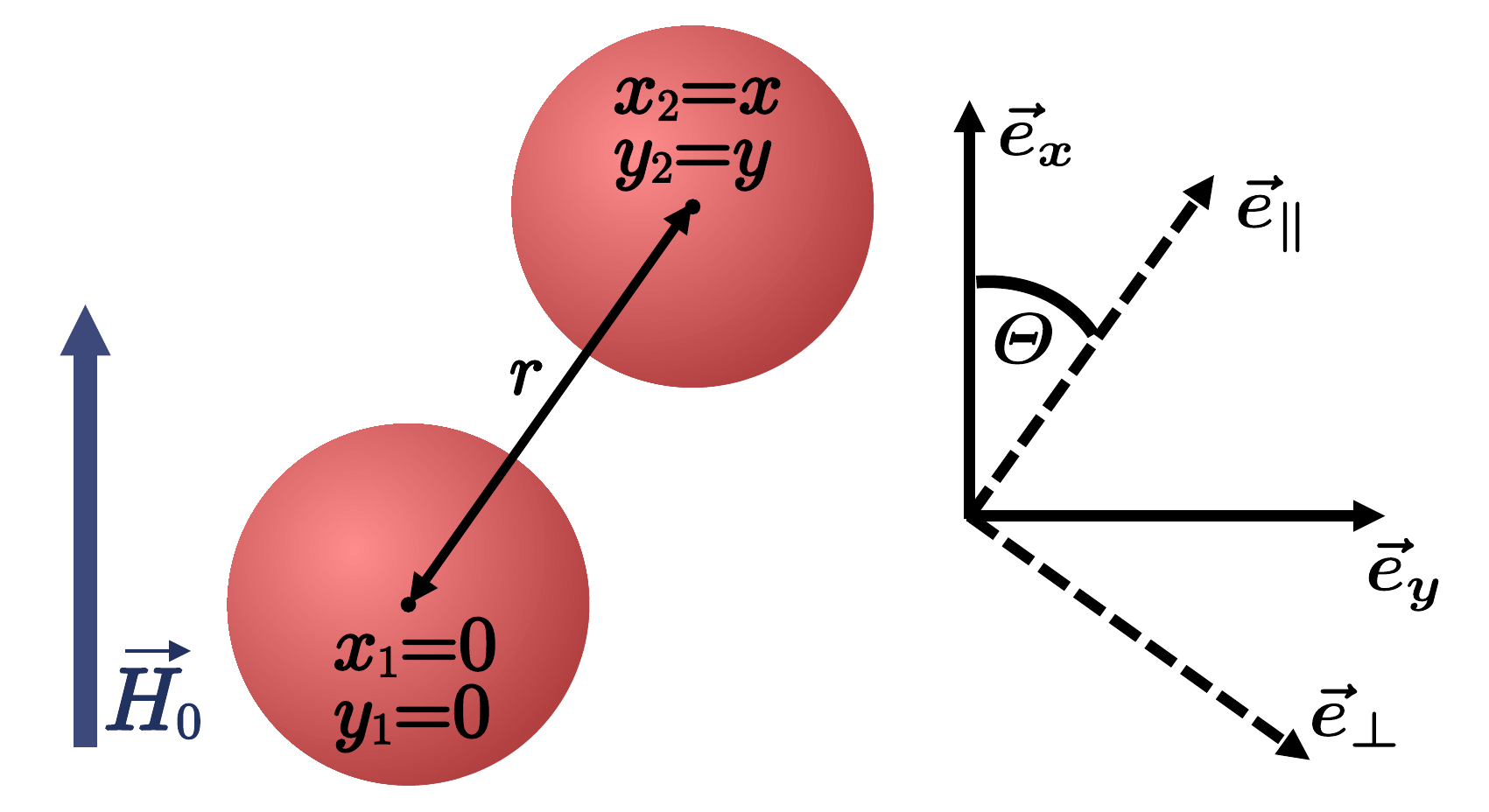}
\caption{\label{fig_Appendix} A sketch of the orientation of the two coordinate systems mentioned in the text.}
\end{figure}

{\it Tensorial form of the solution factor}\,{\bf --}\,In Eq.~(\ref{eq:Mx_2system}), we formulate a solution for the average magnetization along the direction of the applied field, i.e., defined as the $x$-axis. The compact function developed by Biller et al.~\cite{Biller2014, BillerThesis} allows to directly identify the solution factor $\mathcal{F}_{xx}$. To construct the full solution tensor $\hat{\mathcal{F}}$ in Eq.~(\ref{eq:M_2system}), we first note that in the so defined coordinate system, where $\vec{H}_0\parallel\vec{e}_x$, the $x$-$x$-component of $\hat{\mathcal{F}}$ reads ${\mathcal F}_{xx} = 1 + {\mathcal A}(r) + {\mathcal C}(r)\frac{x^{2}}{r^{2}}$. Equivalently, one could also define $\vec{H}_0\parallel\vec{e}_y$, or $\vec{H}_0\parallel\vec{e}_z$, and therefore ${\mathcal F}_{yy} = 1 + {\mathcal A}(r) + {\mathcal C}(r)\frac{y^{2}}{r^{2}}$, resp.~${\mathcal F}_{zz} = 1 + {\mathcal A}(r) + {\mathcal C}(r)\frac{z^{2}}{r^{2}}$. For the same reason, i.e., defining the orientation of $\vec{H}_0$ as either one of the three axis only changes the component labeling, it is clear that $\hat{\mathcal{F}}$ is symmetric with  $\mathcal{F}_{xy}=\mathcal{F}_{yx}$, $\mathcal{F}_{xz}=\mathcal{F}_{zx}$ and $\mathcal{F}_{yz}=\mathcal{F}_{zy}$. To determine these 'cross-components', we examine the same problem in two different coordinate systems, see FIG.~\ref{fig_Appendix} for
$\mathcal{F}_{xy}$. Here, we reduce the problem to two dimensions, e.g.~setting $z=0$. Again we start by defining $\vec{H}_0\parallel\vec{e}_x$ in the first coordinate system. The tensorial form of the solution equation then reads:
\begin{equation}
 \label{eq:A1}
 \langle\vec{M}\rangle=\chi_{\text{eff}}\left(\begin{matrix} \mathcal{F}_{xx} & \mathcal{F}_{xy} \\ \mathcal{F}_{xy} & \mathcal{F}_{yy}\end{matrix}\right)\cdot\vec{H}_0~.
\end{equation}
In the second coordinate system the connection vector $\vec{r}$ between the centers of the two spheres defines one axis, which we denote $\vec{e}_{\parallel}$. The second, perpendicular axis, we name accordingly $\vec{e}_{\perp}$. In this rotated coordinate system, the two spheres are in head-to-tail configuration along $\vec{e}_{\parallel}$ and in side-to-side configuration along $\vec{e}_{\perp}$. Note, that this coordinate system is rotated by an angle $\Theta=\arccos(\frac{x}{r})$ against the first one, as indicated in FIG.~\ref{fig_Appendix}. Accordingly, the solution factor for an external field applied along $\vec{e}_{\parallel}$ would be $\mathcal{F}_{\parallel}= 1 + \mathcal{A}(r) + \mathcal{C}(r)$. If the external field is applied along $\vec{e}_{\perp}$ the solution factor becomes $\mathcal{F}_{\perp}=1 + \mathcal{A}(r)$. Due to symmetry it is clear that when a field is applied along $\vec{e}_{\parallel}$, the average magnetization vector $\langle\vec{M}\rangle$ must also be aligned with $\vec{e}_{\parallel}$. Thus, the perpendicular component must be zero. The same holds for the case that the field is applied along $\vec{e}_{\perp}$, where then $\langle\vec{M}\rangle\parallel\vec{e}_{\perp}$. Therefore, the tensorial form of the solution equation in this second coordinate system is:
\begin{equation}
 \label{eq:A2}
 \langle\vec{M}\rangle=\chi_{\text{eff}}\left(\begin{matrix} {\mathcal F}_{\parallel} & 0 \\ 0 & {\mathcal F}_{\perp}\end{matrix}\right)\cdot\vec{H}_0~.
\end{equation}
For an external field $\vec{H}_0$ aligned along $\vec{e}_x$ we have $\vec{H}_0=H_0\left(\begin{matrix} \cos{\Theta} \\ -\sin{\Theta} \end{matrix}\right)$ in the second, rotated coordinate system. The component $\langle M_y\rangle$ is then found as:
\begin{eqnarray}
\nonumber \langle M_y\rangle = \vec{e}_y\cdot\langle\vec{M}\rangle && = \chi_{\text{eff}}\left(\begin{matrix} \sin{\Theta} \\ \cos{\Theta} \end{matrix}\right)\cdot\left(\begin{matrix} \mathcal{F}_{\parallel}\cos{\Theta} \\ -\mathcal{F}_{\perp}\sin{\Theta} \end{matrix}\right)H_0 \\[0.5em]
 && =\chi_{\text{eff}}\,\mathcal{C}(r)\sin\Theta\cos\Theta\,H_0~.
\end{eqnarray}
With $\Theta=\arccos(\frac{x}{r})$, we have:
\begin{equation}
\mathcal{F}_{xy} = \mathcal{C}(r)\frac{xy}{r^{2}}~.
\end{equation}
Equivalently, also the factors $\mathcal{F}_{xz}$ and $\mathcal{F}_{yz}$ are found, and the solution tensor $\hat{\mathcal{F}}$ takes the form presented in Eq.~(\ref{eq:M_2system}). 

{\it Analytic expression for the magnetic full-field forces}\,{\bf --}\,In the case of dipole interactions, the forces can be computed directly using the classical expression \cite{Biller2015b, Metsch2021}. Since we now have an analogue operator form, the same methodology can also be employed in the case of full-field interactions. Generally we have \cite{Metsch2021}:
\begin{equation}
 \label{eq:magnetic_force}
 \vec{f}_i = \mu_0 v_p \sum_{j\neq i}^{N}\left\lbrace \langle\vec{M}_i\rangle\cdot\left(\,\vec{\nabla}_i \hat{\mathcal G}_{ij}\right)\cdot\langle\vec{M}_j\rangle\right\rbrace
\end{equation}
Thanks to the given form of $\hat{\mathcal{G}}$ the gradient $\vec{\nabla}\hat{\mathcal G}(\vec{r})$ is obtained in analytic form resulting in a 3rd order tensor:
\begin{eqnarray}
\label{eq:gradG}
\nonumber \vec{\nabla}\hat{\mathcal G} = {\mathcal G}'_{I}(r)\sum\limits_{a=x}^{z}\vec{e}_a\frac{\vec{r}}{r}\vec{e}_a  + \frac{{\mathcal G}_{R}(r)}{r}\left( \hat{I}\frac{\vec{r}}{r} + \frac{\vec{r}}{r}\hat{I}\right) \\ + \left({\mathcal G}'_{R}(r)-2\frac{{\mathcal G}_{R}(r)}{r}\right)\frac{\vec{r}\vec{r}\vec{r}}{r^{3}}~.
\end{eqnarray}
Note that using the dipole coefficients $g_I\to\mathcal{G}_I$ and $g_R\to\mathcal{G}_R$ in Eqs.~(\ref{eq:magnetic_force}) and (\ref{eq:gradG}) also leads to the classical dipole forces. Multiplying the gradient of $\hat{\mathcal{G}}$ from the left and the right by the corresponding magnetization vectors according to Eq.~(\ref{eq:magnetic_force}), the force is finally computed as:
\begin{eqnarray}
\label{eq:FF_force}
\nonumber \vec{f}_i = -\mu_0 v_p \sum_{j\neq i}^{N} && \left\lbrace {\mathcal G}'_{I}(r_{ij}) \left(\langle\vec{M}_j\rangle\cdot\langle\vec{M}_i\rangle\right)\frac{\vec{r}_{ij}}{r_{ij}}\right.\\
 \nonumber &&+\, \frac{{\mathcal G}_{R}(r_{ij})}{r_{ij}}\bigg(\!\!\left(\langle\vec{M_j}\rangle\cdot\frac{\vec{r}_{ij}}{r_{ij}}\right)\langle\vec{M}_i\rangle + \left(\langle\vec{M_i}\rangle\cdot\frac{\vec{r}_{ij}}{r_{ij}}\right)\langle\vec{M}_j\rangle \bigg) \\
 && \left. + \left({\mathcal G}'_{R}(r_{ij})-2\frac{{\mathcal G}_{R}(r_{ij})}{r_{ij}}\right)\!\left(\langle\vec{M}_j\rangle\cdot\frac{\vec{r}_{ij}}{r_{ij}}\right)\left(\langle\vec{M}_i\rangle\cdot\frac{\vec{r}_{ij}}{r_{ij}}\right)\frac{\vec{r}_{ij}}{r_{ij}} \right\rbrace\,.
\end{eqnarray}
Thus, it is not necessary to take the detour via the total magnetic energy in Eq.~(\ref{eq:Umag}) and find the forces by computing the  gradient numerically. Instead, the self-consistent $\langle\vec{M}_i\rangle$ can be directly plugged into Eq.~(\ref{eq:FF_force}). This provides an additional computational advantage of the present full-field operator form. 

\end{document}